\documentclass[preprints,article,accept,moreauthors,pdftex]{Definitions/mdpi} 

\firstpage{1} 
\pubvolume{xx}
\issuenum{1}
\articlenumber{5}
\pubyear{2019}
\copyrightyear{2019}
\history{Received: date; Accepted: date; Published: date}

\pdfoutput=1

\usepackage{hyperref}

\Title{Correlation of heavy and light flavours in simulations}

\Author{Eszter Frajna $^{1,2}$ \orcidA{}, Róbert Vértesi$^{2}$\orcidB{}}

\AuthorNames{Frajna Eszter, Vértesi Róbert}

\address{%
$^{1}$ \quad Budapest University of Technology and Economics\\
$^{2}$ \quad Wigner Research Centre for Physics$^{\dagger}$}

\corres{Correspondence: frajna.eszter@wigner.mta.hu}
\firstnote{Current address: 1121 Budapest, Konkoly-Thege Miklós út 29-33, Hungary}

\abstract {The ALICE experiment at the Large Hadron Collider (LHC) ring is designed to study the strongly interacting matter at extreme energy densities created in high-energy heavy-ion collisions. In this paper we investigate correlations of heavy and light flavours in simulations at LHC energies at mid-rapidity, with the primary purpose of proposing experimental application of these methods. Our studies have shown that investigating the correlation images can aid the experimental separation of heavy quarks and help understanding the physics that create them. The shape of the correlation peaks can be used to separate the electrons stemming from b quarks. This could be a method of identification that, combined with identification in silicon vertex detectors, may provide much better sample purity for examining the secondary vertex shift. Based on a correlation picture it is also possible to distinguish between prompt and late contributions to D meson yields.}


\begin{document}

\section{Introduction}
Quark–gluon plasma (QGP) is a state of matter which exists at extremely high temperatures and densities, where quarks are no longer confined to hadrons \cite{Shuryak:1980tp}. Initially, the Universe was filled with this hot and dense matter. Quark-gluon plasma can be recreated in high-energy heavy-ion collisions at large accelerator rings such as the LHC. 

The angular correlation data of the STAR experiment \cite{Adams:2003im} shows a strong suppression of the away-side correlation peak in central Au+Au collisions,  while no such effect can be observed in p+p collisions. This indicates a strong quenching of jets that traverse the QGP. In the case of d+Au collision, we do not experience jet-suppression even though there is cold hadronic nuclear matter present. It was among the first convincing evidences of hot and dense strongly interacting nuclear matter in the final state. The interaction of partons with Quark-gluon plasma is often studied by full jet reconstruction. However, in heavy-ion collisions, high background from the underlying event makes it difficult to reconstruct jets below a certain momentum. Measuring the angular correlation of particles is a technique that solves this problem. Comparison of angular correlations in small and large collision systems reveals information of jet modification by the strongly interacting medium.

Heavy-flavour (charm and beauty) quarks are an excellent tool to study heavy-ion collisions. Most of them are created in the initial stages of the reaction, and their long lifetime ensures that they interact both with the hot and dense medium as well as with the cold hadronic matter before they decay. Identifying characteristic correlation images of heavy and light quarks can help understand flavour-dependent fragmentation. Furthermore, finding these characteristic shapes can be used as an alternative method to pin down feed-down of beauty hadrons into charm without the need to find the secondary vertex. Comparing the near-side and away-side  correlation peaks associated with hadrons from different heavy quarks from a given $p_T$. The following simulation results serve as a case study that the techniques we developed can be successfully applied in today's large experiments such as ALICE at the LHC to aid and interpret heavy-flavor correlation measurements. 

\section{Analysis method}
The variables we use to describe particle kinematics are the three-momentum ($p_x, p_y, p_z$), the azimuth angle ($\varphi=\frac{p_y}{p_x}$) and the pseudorapidity ($ \eta = \frac{1}{2} \mathrm{ln} \frac{p+p_\mathrm{z}}{p-p_\mathrm{z}} $, where $ p_T=\sqrt{{p_x}^2+{p_y}^2}$ is the transverse momentum).

To examine the correlation of particles, we followed a technique similar to \cite{Adam:2016tsv}. We selected a trigger particle from a given momentum ($p_T$) range, and associated particles from a lower momentum ($p_T$) window to examine all other particles from the same event. Then we calculated the differences in the azimuth angles as well as the pseudorapidities of the trigger and associated particles within each event. The plane ($\Delta \eta, \Delta \varphi$) can be divided into two parts along $\Delta \varphi$: near-side and away-side peak region. We define near-side as the range from -$\pi/2$ to $\pi/2$, and the away-side as the range from $\pi$/2 to 3$\pi$/2. Near-side correlations give information about the structure of jets, since after background subtraction, most trigger and associated particles come from the same jet. Differences between correlations of beauty, charm and light flavour provide valuable information about flavour-dependent jet fragmentation. Away-side correlation is mostly from back-to-back jet pairs and it is sensitive to the underlying hard processes.

We examined the width of fitted functions in the $5<p_\mathrm{T}^{trigger}<8$ GeV/$c$ trigger particle transverse momentum range, and in different associated particle transverse momentum ($p_\mathrm{T}^{assoc}$) ranges (${1<p_\mathrm{T}^{assoc}<2}$ GeV/$c$, $2<p_\mathrm{T}^{assoc}<3$ GeV/$c$, $3<p_\mathrm{T}^{assoc}<5$ GeV/$c$ and $5<p_\mathrm{T}^{assoc}<8$ GeV/$c$). In the latter case we ensured that $p_\mathrm{T}^{trigger}$ > $p_\mathrm{T}^{assoc}$ . 

Correlation analyses usually use the mixed-event technique to correct for the finite size of the detector in $\eta$ \cite{Adam:2016tsv}. However, at mid-rapidity it is sufficient to assume a uniform track distribution and therefore we apply a reweighting of events with a "tent-shaped" function $\frac{dN}{\Delta \eta} = \frac{1}{2A} - \frac{|\eta|}{4A^2}$ where $A=2$ is the maximal acceptance in which tracks were recorded. In the analysis, we only used the |$\Delta \eta$|<1.6 range.


We used the PYTHIA 8.1 Monte-Carlo event generator \cite{Sjostrand:2007gs} to simulate hard QCD events using the default Monash 2013 \cite{Monash 2013} settings for LHC p+p data. The phase space has been reduced so that the leading hard process has at least 5 GeV/$c$ momentum. Heavy-flavour scale settings were used similarly to recent STAR analyses, eg. that in Ref.~\cite{Adamczyk:2016dzv}.


Processes that fulfill the conditions of the Central Limit Theorem yield Gaussian-shaped correlation peaks. In case of long-lived resonances, however, the Gaussian shape is not necessarily adequate. Therefore we also apply a Generalized Gaussian fit on the near- and away-side correlation peaks. 
Gaussian functions of the different projections are 
\begin{equation}
f(\Delta\varphi)=N \cdot \frac{1}{\sqrt{2 \pi} \sigma_{\Delta \varphi}} \cdot e^{-(\frac{\Delta \varphi ^2}{2 \sigma_{\Delta \varphi}^2})} ,~~ f(\Delta\eta)= N \cdot \frac{1}{\sqrt{2 \pi} \sigma_{\Delta\eta}} \cdot e^{-(\frac{\Delta \eta ^2}{2 \sigma_{\Delta\eta}^2})},
\end{equation}
and the generalized Gaussian functions of the different projections are 
\begin{equation}
g(\Delta\varphi)= N \cdot \frac{\gamma_{\Delta \varphi}} {2\omega_{\Delta \varphi} \Gamma(\frac{1}{\gamma_{\Delta\varphi}})} \cdot e^{-(\frac{|\Delta \varphi|}{\omega_{\Delta\varphi}})^{\gamma_{\Delta\varphi}}},~~g(\Delta\eta)=N \cdot \frac{\gamma_{\Delta\eta}} {2\omega_{\Delta\eta} \Gamma(\frac{1}{\gamma_{\Delta\eta}})} \cdot e^{-(\frac{|\Delta \eta|}{\omega_{\Delta\eta}})^{\gamma_{\Delta\eta}}},
\end{equation}
where N is the normalization factor, $\sigma_{\Delta\varphi}$ is the width of the peak in the direction of $\Delta \varphi$, and $\sigma_{\Delta \eta}$ is the width of the peak in the direction of $\Delta \eta$.

The generalized Gaussian function has an extra parameter $\gamma$ compared to the Gaussian. If $\gamma$ = 1, then the generalized Gaussian function is an exponential function. If $\gamma$ = 2, it is reduced to a regular Gaussian function. And if $\gamma$ is greater than two then the top of the function is flattened.

\section{Results}
\subsection{Correlations of light charged hadrons}
As a first test we reproduce the near- and away-side correlation peaks of light charged hadrons ($\pi^\pm$, K$^\pm$, p and $\bar{\mathrm{p}}$) both in the $\Delta\eta$ and the $\Delta\varphi$ directions. For all the particles we investigate in this study, we only examine the near-side peak in the direction of $\Delta \eta$, and in the direction of $\Delta \varphi$ we investigate both peaks. Below, the parameters are shown in different $p_\mathrm{T}^{assoc}$ ranges.

\begin{figure}[H]
	\centering
	{{\includegraphics[width=7cm]{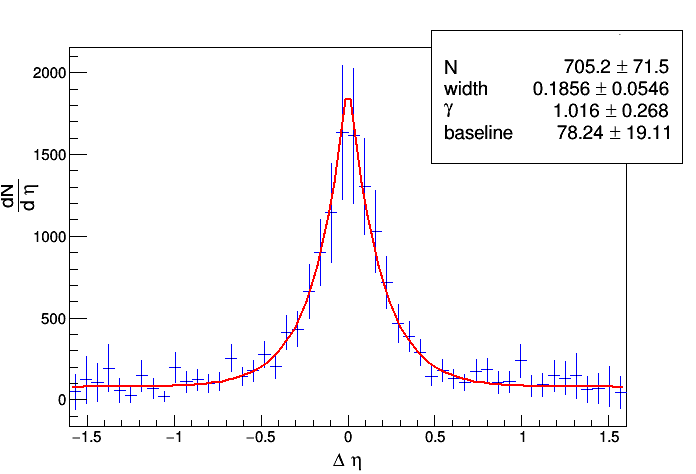} }}%
	\qquad
	{{\includegraphics[width=7cm]{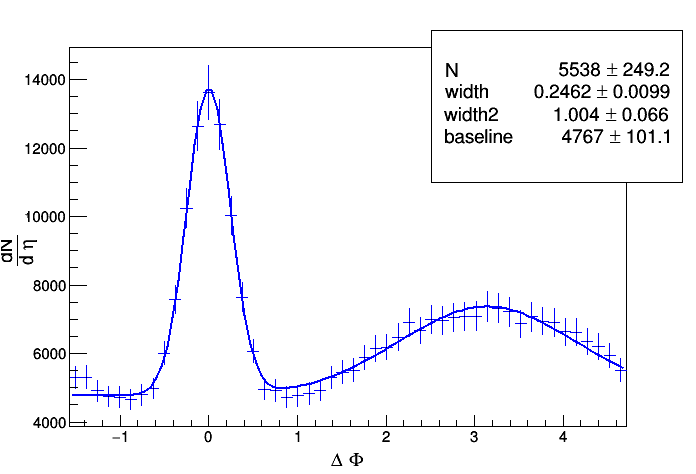} }}%
	\caption{The near- and away-side peaks of light charged hadrons with Gaussian fitting (on the left side) and with generalized Gaussian fitting (on the right side). }%
	\label{fig:1}%
\end{figure}

The left panel in Fig.~\ref{fig:1} represents the near-side peak with generalized Gaussian fit. The parameter $\gamma$ is unity within uncertainties, indicating a distribution that is significantly sharper than Gaussian and consistent with an exponential function. The right panel in Fig.~\ref{fig:1} shows the near- and away-side peaks with Gaussian fits in $\Delta \varphi $ direction. The shape of the peaks are well described by a Gaussian.

\begin{figure}[H]
	\centering
	{{\includegraphics[width=7cm]{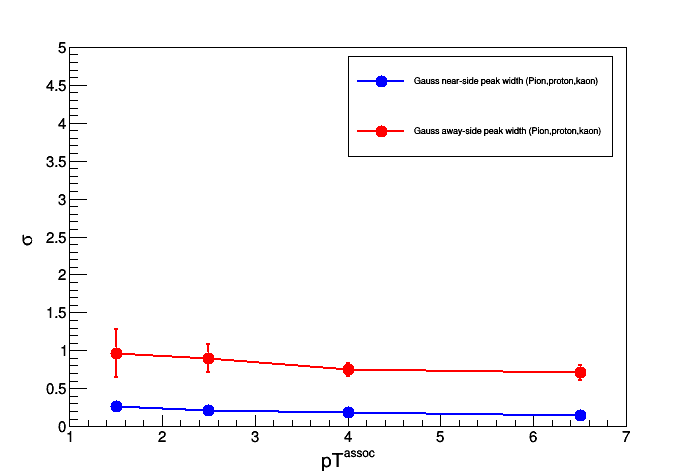} }}%
	\qquad
	{{\includegraphics[width=7cm]{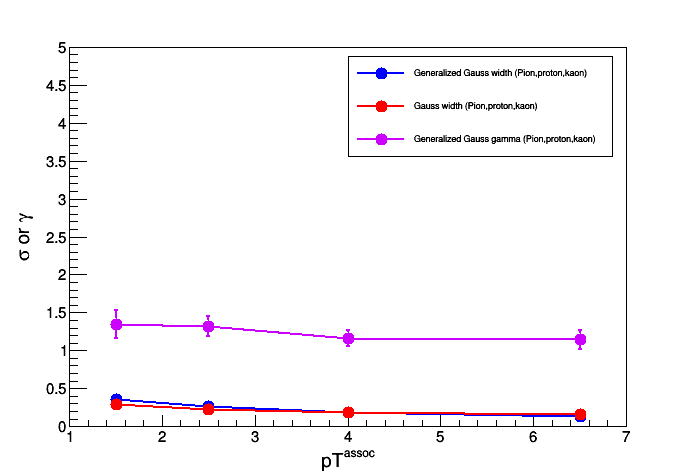} }}%
	\caption{The peak width for Gaussian fitting (on the left side) and for generalized Gaussian fitting (on the right side) for 5<$p_\mathrm{T}^{trigger}$<8 GeV/$c$ and different $p_\mathrm{T}^{assoc}$ values.}%
	\label{fig:3}%
\end{figure}

To get a more comprehensive picture, we also show the fit parameters in function of $p_T$. The left panel in Fig.~\ref{fig:3} shows the peak width for Gaussian fitting, while the right panel in Fig.~\ref{fig:3} shows the peak width for generalized Gaussian fitting and the $\gamma$ parameter of the function. The correlation peaks of the light charged hadrons are getting narrower towards higher $p_T$ and $\gamma$ is constant within uncertainties.

\subsection{Prompt production of heavy flavour mesons}
\begin{figure}[H]
	\centering
	{{\includegraphics[width=7cm]{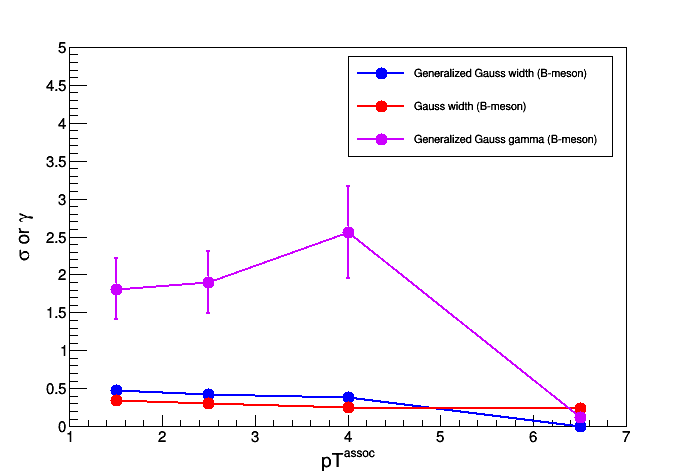} }}%
	\qquad
	{{\includegraphics[width=7cm]{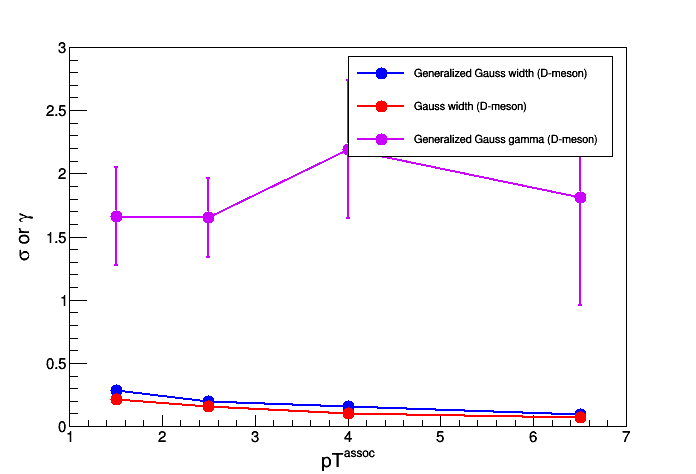} }}%
	\caption{D mesons from c quarks (on the left side) and B mesons from b quarks(on the right side), for $5<p_\mathrm{T}^{trigger}<8$ GeV/$c$ and different $p_\mathrm{T}^{assoc}$ values.}%
	\label{fig:5}%
\end{figure}

We examined the direct decay cases of D mesons from c quarks and B mesons from b quarks, without feed-down. In Fig.~\ref{fig:5} both peaks are consistent with Gaussian. The last point in the left panel in Fig. \ref{fig:5} is due to a non-convergent fit.

\subsection{D meson from the decay of the B meson}
\begin{figure}[H]
	\centering
	{{\includegraphics[width=7cm]{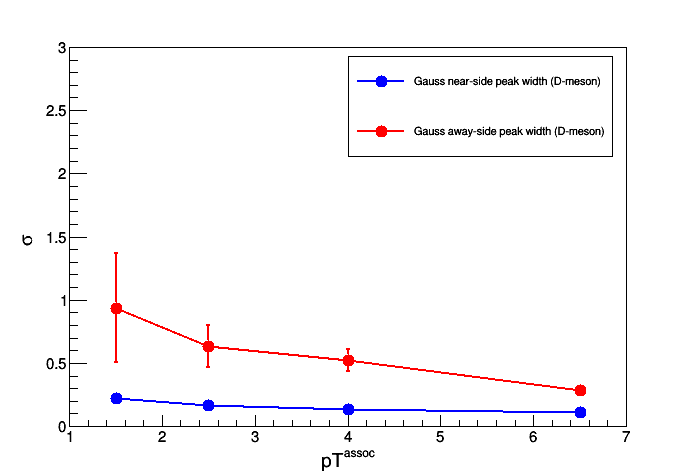} }}%
	\qquad
	{{\includegraphics[width=7cm]{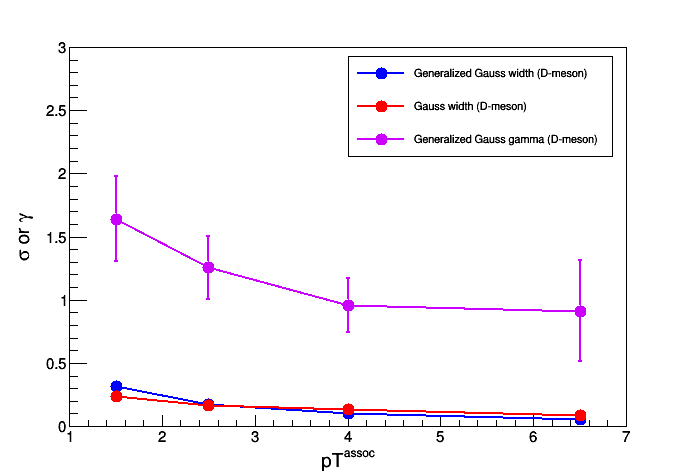} }}%
	\caption{D meson from the decay of the B meson, for $5<p_\mathrm{T}^{trigger}<8$ GeV/$c$ and different $p_\mathrm{T}^{assoc}$ values.}%
	\label{fig:7}%
\end{figure}

We investigated D mesons from the decay of B mesons. The left panel in Fig.~\ref{fig:7} shows a strong dependence of the away-side peak on $p_T$ (not observed in light flavor or prompt heavy flavour production).
The right panel in Fig.~\ref{fig:7} $\gamma$  decreases with $p_T$, together with $\gamma$. (Peaks are getting both narrower and $\gamma$ is less than two towards high $p_T$).

Based on Figs.~\ref{fig:5} and ~\ref{fig:7}, it is possible to distinguish between prompt D mesons and non-prompt D mesons from decays of B mesons. The statistical separation of these two contributions allows for the understanding of the flavor-dependence of heavy-quark energy loss within the Quark-gluon plasma. The effect of Quark-gluon plasma on heavy quarks can be investigated by the ratio of D meson from b quarks and c quarks.

\subsection{Investigation of electrons from B mesons}
We investigated the electrons from B mesons. Electrons can come directly from semileptonic B-decays, as well as semileptonic decays of charmed mesons that B feeds down into. Still, these branching ratios are in the order of a couple of percents \cite{PDG}, so the electron yield is relatively low. Therefore, the lack of statistics was a limit in the analysis. However, the results are suitable for drawing conclusions.

\begin{figure}[H]
	\centering
	{{\includegraphics[width=7cm]{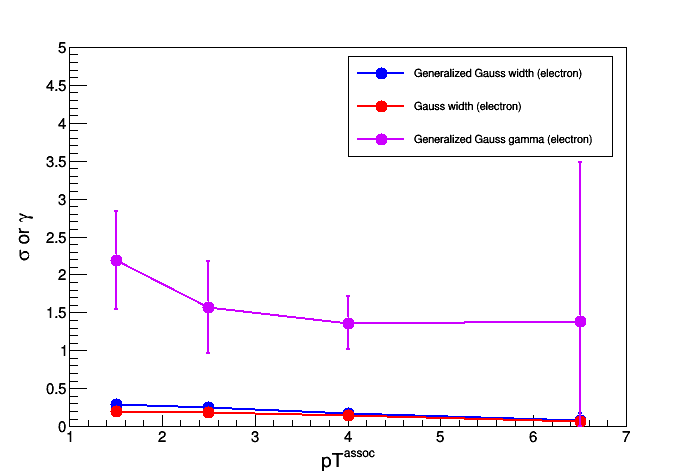} }}%
	\qquad
	{{\includegraphics[width=7cm]{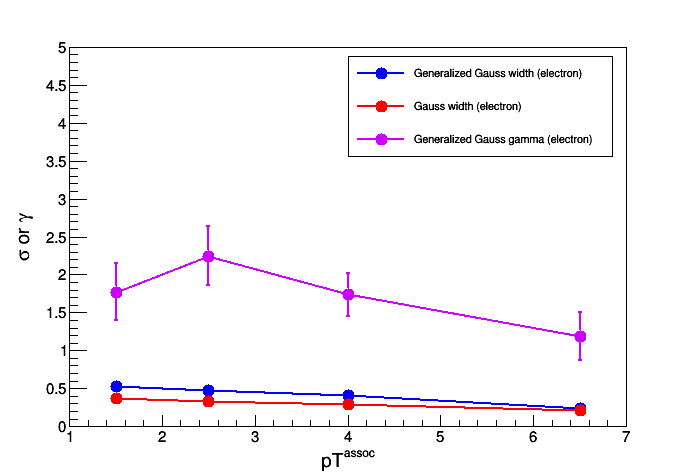} }}%
	\caption{Generalized Gaussian fitting function for electrons from B mesons (left) and c quarks (right), for $5<p_\mathrm{T}^{trigger}<8$ GeV/$c$ and different $p_\mathrm{T}^{assoc}$ values.}%
	\label{fig:9}%
\end{figure}

Fig.~\ref{fig:9} shows that correlations of B meson decay electron with hadron produce wider correlation peaks than in the c quark decay electron case. There's no significant dependence of $\gamma$ on $p_\mathrm{T}^{assoc}$ and $\gamma \sim$2.

It can be assumed that the significant deviation of the peaks from Gaussian and the strong momentum dependence of the parameters can be traced back to the decay kinematics, in which the momentum of the b quark and the location of the secondary vertex play a role. It is a well-known phenomenon that long-range correlation components from long-lived resonances lead to Lévy-like distributions rather than Gaussian~\cite{Vertesi:2007ki}. It seems possible to separate the electrons coming from B mesons simply by the shape of the correlation peaks. This could be a method of identification that, combined with particle identification based on secondary vertex reconstruction in silicon tracking detectors such as the ALICE ITS \cite{Yang:2015yia},  may provide a much better sample purity than what we can currently achieve.

\subsection{Comparison of B-meson and b-quark correlations}
Finally, we compared correlations of B mesons with hadrons to b quarks with hadrons. The b quarks were taken directly from parton-level Monte Carlo truth information. In an experiment, correlations with b quarks can be constructed by taking the jet axis of b-tagged jets ~\cite{arXiv:1410.2576}, a process that is very problematic at low momenta, especially in heavy ion collisions. We expect that there will be no significant difference between the two, as the only source of B meson is the b quark decay, besides the b quark direction is close to the axis of the b-jet, and due to its large mass the direction of the b quark momentum will determine the direction of the B meson momentum. However, higher precision is needed to verify whether the away-side peak of b quark to hadron correlations follows a similar trend to the away-side peak of B meson to hadron correlations.

\begin{figure}[H]
	\centering
	{{\includegraphics[width=7cm]{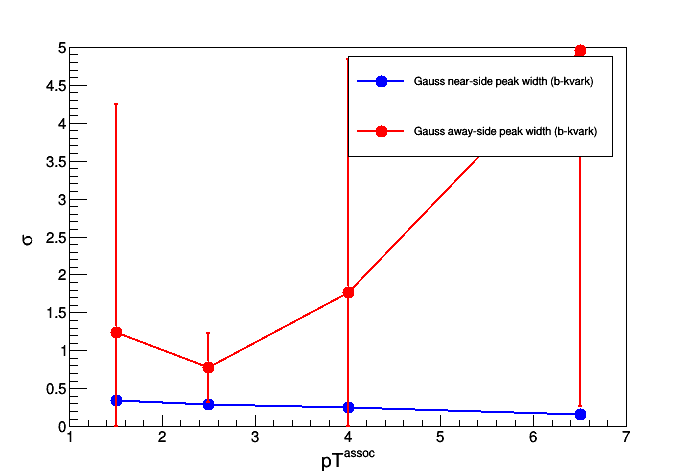} }}%
	\qquad
	{{\includegraphics[width=7cm]{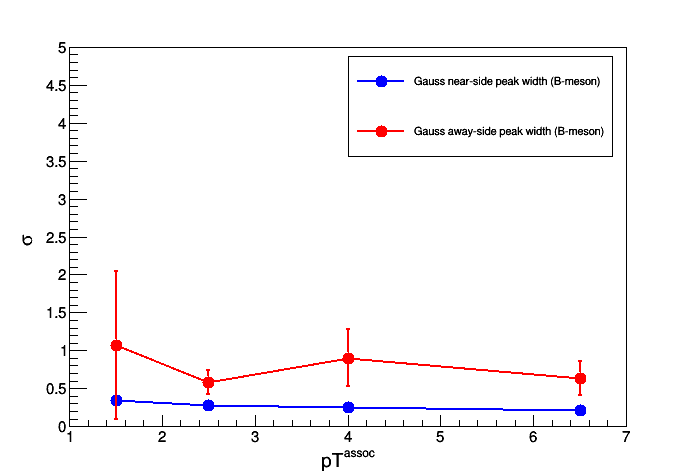} }}%
	\caption{Comparsion of b quark (on the left) and B meson (on the right) correlations to hadrons.}%
	\label{fig:11}%
\end{figure}

Evolution of correlation pictures with momentum match within uncertainties. We have a characteristic b correlation image, which is present in both b quarks and B mesons, which is further support that the B meson is a good proxy for the b quark.

\section{Conclusions}

In summary, we have shown that a detailed analysis of heavy-flavor correlations can help in understanding flavor-dependent fragmentation as well as aid particle identification. The shape of the correlation peaks can be used to separate the  electrons coming from b quark decays. This could be a method of identification that, combined with particle identification in secondary vertex detectors, may provide a much better sample purity than traditional methods. Correlation images are sensitive to the distribution of secondary vertices of heavy-quark decays, and the latter processes can be statistically separated from light quarks. Based on a correlation picture it is possible to distinguish between prompt D mesons and non-prompt D mesons from decays of B mesons. The statistical separation of these two contributions allows for the understanding of the flavor-dependence of heavy-quark energy loss within the quark-gluon plasma. We also see a characteristic b-correlation image, which is present in both b quarks and B mesons. B mesons can be used to study b quarks in the momentum regime where the reconstruction of b jets is not feasible.

\vspace{6pt} 



\authorcontributions{This research was carried out by E.F. under the supervision of R.V. }

\funding{This work has been supported by the Hungarian National Research Fund
(OTKA) grant K120660. Author RV thanks for the support of the J\'anos Bolyai fellowship of the Hungarian Academy of Sciences.}

\conflictsofinterest{The author declares no conflict of interest.}




\reftitle{References}



\sampleavailability{Data are available from the authors on request.}


\end{document}